\documentclass[14pt]{elsart}
\usepackage{refmerge}
\usepackage{epsfig}
\begin{document}
\begin{frontmatter}
\date{}
\title{\bf{ \boldmath
TOTAL CROSS SECTION OF THE PROCESS $e^+e^-\to\pi^+\pi^-\pi^+\pi^-$
IN THE C.M.ENERGY RANGE 980--1380 MEV
}}
%
%

\author[BINP]{R.R.~Akhmetshin},
\author[BINP,NGU]{V.M.~Aulchenko},
\author[BINP]{V.Sh.~Banzarov},
\author[PITT]{A.~Baratt},
\author[BINP,NGU]{L.M.~Barkov},
\author[BINP]{N.S.~Bashtovoy},
\author[BINP,NGU]{A.E.~Bondar},
\author[BINP]{D.V.~Bondarev},
\author[BINP]{A.V.~Bragin},
\author[BINP,NGU]{S.I.~Eidelman},
\author[BINP]{D.A.~Epifanov},
\author[BINP,NGU]{G.V.~Fedotovitch},
\author[BINP]{N.I.~Gabyshev},
\author[BINP]{D.A.~Gorbachev},
\author[BINP]{A.A.~Grebeniuk}, 
\author[BINP,NGU]{D.N.~Grigoriev},
\author[BINP]{F.V.~Ignatov},
\author[BINP]{S.V.~Karpov},
\author[BINP,NGU]{V.F.~Kazanin},
\author[BINP,NGU]{B.I.~Khazin},
\author[BINP,NGU]{I.A.~Koop},
\author[BINP,NGU]{P.P.~Krokovny},
\author[BINP,NGU]{A.S.~Kuzmin},
\author[BINP,BOST]{I.B.~Logashenko},
\author[BINP]{P.A.~Lukin},
\author[BINP]{A.P.~Lysenko},
\author[BINP]{K.Yu.~Mikhailov},
\author[BINP,NGU]{A.I.~Milstein},
\author[BINP,NGU]{I.N.~Nesterenko},
\author[BINP]{V.S.~Okhapkin},
\author[BINP]{A.V.~Otboev},
\author[BINP,NGU]{E.A.~Perevedentsev},
\author[BINP]{A.A.~Polunin},
\author[BINP]{A.S.~Popov},
\author[BINP]{S.I.~Redin},
\author[BINP]{N.I.~Root},
\author[BINP]{A.A.~Ruban},
\author[BINP]{N.M.~Ryskulov},
\author[BINP]{A.G.~Shamov}, 
\author[BINP]{Yu.M.~Shatunov},
\author[BINP,NGU]{B.A.~Shwartz},
\author[BINP]{A.L.~Sibidanov},
\author[BINP]{V.A.~Sidorov}, 
\author[BINP]{A.N.~Skrinsky},
\author[BINP]{I.G.~Snopkov},
\author[BINP,NGU]{E.P.~Solodov},
\author[PITT]{J.A.~Thompson}, 
\author[BINP]{A.A.~Valishev},
\author[BINP]{Yu.V.~Yudin},
\author[BINP,NGU]{A.S.~Zaitsev},
\author[BINP]{S.G.~Zverev}

\address[BINP]{Budker Institute of Nuclear Physics, 
  Novosibirsk, 630090, Russia}
\address[BOST]{Boston University, Boston, MA 02215, USA}
\address[NGU]{Novosibirsk State University, 
  Novosibirsk, 630090, Russia}
\address[PITT]{University of Pittsburgh, Pittsburgh, PA 15260, USA}

%
%
\vspace{0.7cm}
\begin{abstract}
\hspace*{\parindent}
The $e^+e^-\to\pi^+\pi^-\pi^+\pi^-$ cross section has been measured
using 5.8 pb$^{-1}$ of integrated luminosity collected with the CMD-2
detector at the VEPP-2M collider. 
Analysis of the data with a refined efficiency determination
and use of both three- and four-track events  
allowed doubling of a data sample and reduction of systematic errors 
to 5-7\%.

\end{abstract}
\end{frontmatter}
\maketitle
\baselineskip=17pt
\section*{ \boldmath Introduction}
\hspace*{\parindent}
Production of four pions in $e^+e^-$ annihilation 
is the dominant process contributing to the total hadronic cross 
section in the c.m. energy range between 1000 and 2000~MeV. Precise
measurements of the cross sections of the reactions
$e^+e^- \to 2\pi^+2\pi^-, \pi^+\pi^-2\pi^0$ 
will improve the accuracy of the calculation 
of the hadronic contribution to the muon anomalous magnetic 
moment~\cite{g-2,cvc1} and provide an important input 
to tests of the relation between the cross sections of the process 
$e^+e^- \to 4\pi$ and the differential rate of the 
$\tau^{\pm} \to (4\pi)^{\pm} \nu_{\tau}$ decay following from
the conservation of the vector current and isospin symmetry~\cite{cvc1,cvc2}.
As one of the possible decay modes of the isovector vector states,
a four-pion final state and various mechanisms of its production 
can provide information on the properties of the $\rho$ excitations
as well as shed light on the problem of existence  of light exotic
states (hybrids) between 1000 and 2000~MeV~\cite{hybrid1,hybrid2}.

Although the process $e^+e^- \to 2\pi^+2\pi^-$ has been
extensively studied before in the c.m.energy range 1000--1400~MeV 
by various groups at the VEPP-2M collider
in Novosibirsk~\cite{olya,cmd,NDR,CMD2_4pi,SND_4pi2},  
the scatter of the obtained results as well as their systematic 
uncertainties are rather big. 
In the previous analysis of this process at CMD-2, which was  focused 
on its dynamics, we reported on the first observation of the 
$a_{1}(1260)\pi$ dominance~\cite{CMD2_4pi}.
Later this result was confirmed by the CLEO~\cite{CLEO_4pi} and 
SND~\cite{SND_4pi2} groups.  
In this paper we present a new analysis of the same data sample based on 
5.8 pb$^{-1}$ of integrated luminosity collected at CMD-2
at 36 energy points in the 980-1380 MeV range  with a 10 MeV step. 
A new reconstruction algorithm combined with refined 
detector calibrations 
and an update of the integrated luminosity~\cite{update} as well as 
use of both three- and four-track events for the cross section
determination allowed a new measurement of the cross section with
smaller statistical and systematic uncertainties. 
The values of the cross section obtained in this work supersede
our previous results in~\cite{CMD2_4pi}.

The general purpose detector CMD-2 has been described in 
detail elsewhere~\cite{cmddet}. Its tracking system consists of a 
cylindrical drift chamber (DC) and double-layer multiwire proportional 
Z-chamber, both also used for a trigger, and both inside a thin 
(0.38~X$_0$) superconducting solenoid with a field of 1~T. 
The barrel CsI calorimeter with a thickness of 8.1~X$_0$ is placed
outside  the solenoid  and the end-cap BGO calorimeter with a 
thickness of 13.4~X$_0$ is placed inside the solenoid.
The luminosity is measured using events of Bhabha scattering 
at large angles~\cite{prep}. 

%
\section*{Selection of $e^+e^-\to\pi^+\pi^-\pi^+\pi^-$ Events}
\hspace*{\parindent}
Candidates for the process under study were selected from a data sample 
containing three and more charged tracks reconstructed in the DC 
and possessing the following properties:

\begin{itemize}
\item{
A track contains more than six points in the R-$\phi$ plane.
}
\item{
A track momentum does not exceed a beam
momentum by more than 10\%.
}
\item{
A minimum distance from the track to the beam axis in the
R-$\phi$  plane is less than 0.5 cm.
}
\item{
A minimum distance from the track to the center of the interaction
region  along Z  is less than 10 cm.
}
\item{
A track has a polar angle $\theta$ big enough to cross half of the 
DC radius and produce enough hits of the DC wires for a good track
reconstruction.
}
\end{itemize}

Events with three and four tracks satisfying the above  requirements
were considered as candidates
for the $e^+e^-\to\pi^+\pi^-\pi^+\pi^-$ process. 
About 26200 four-track events and 22800 three-track events were selected. 
The number of events with
five or more selected  tracks was found to be negligible.
\begin{center}
\begin{figure}[tbh]
\psfig{file=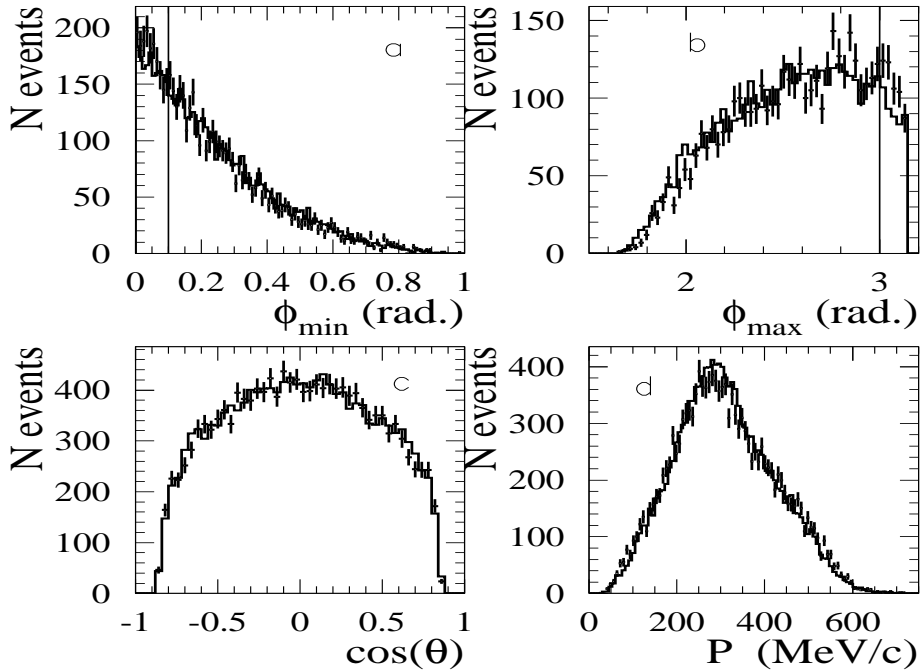, 
width=0.8\textwidth, height=0.6\textwidth}
\caption
{
Distributions for four-track experimental events (points with errors)
and simulation (histograms) at 2E$_{\rm beam}$=1380~MeV:
(a) Minimum angle between two tracks in the R-$\phi$ plane;
(b) Maximum angle between two tracks  in the R-$\phi$ plane;
(c) Cosine of the track polar angle;
(d) Track momentum. The lines show applied cuts.
}
\label{distr1}
\end{figure}
\end{center}

\hspace*{\parindent}Reconstructed momenta 
and angles of the tracks for four-track events were used for further 
selection. Figure~\ref{distr1} presents various distributions
for selected events at 2E$_{\rm beam}$=1380~MeV.

The following cuts are additionally applied to further suppress 
background events. A requirement for a minimum angle between two 
tracks in the R-$\phi$ plane to be greater than 0.1 radian removes  
background events from the processes $e^+e^-\to\pi^+\pi^-\pi^0(\pi^0)$
with photon conversion to an $e^+e^-$ pair, see Fig.~\ref{distr1}(a).
A requirement for a maximum  angle between two tracks  in the 
R-$\phi$ plane to be less than 3.0 radian suppresses background from the 
$K^+K^-$ pair production (kaons have a high probability to decay 
inside the DC and produce additional tracks) and cosmic showers,
see Fig.~\ref{distr1}(b). Figures~\ref{distr1}(c),(d) present the 
$cos(\theta)$ and momentum distributions for detected tracks after 
applying cuts on relative angles. 
Results of the Monte Carlo simulation (MC) presented by open histograms 
well describe the kinematical parameters in Figs.~\ref{distr1}(a)-(d).



\begin{center}
\begin{figure}[tbh]
\vspace{-0.2cm}
\psfig{file=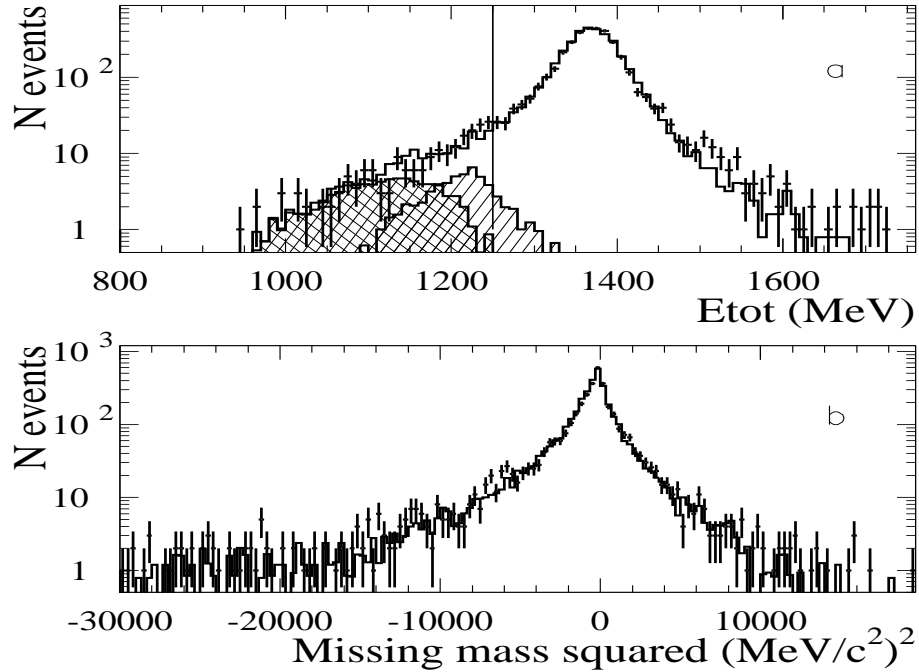, 
width=0.8\textwidth, height=0.6\textwidth}
\caption
{
Distributions for four-track experimental events (points with errors)
and simulation (open histograms) at 2E$_{\rm beam}=$~1380 MeV:
(a) The total energy of four pions. The hatched and cross-hatched 
histograms show the contributions from five-pion production 
($\omega\pi^+\pi^-$ and $\eta\pi^+\pi^-$, respectively). The line shows 
an applied cut; (b) Missing mass squared for four pions. 
}
\label{distr2}
\end{figure}
\end{center}

Figure~\ref{distr2}(a) presents the total energy distribution for events 
with four  tracks after the above selections at 
2E$_{\rm beam}=$~1380 MeV. 
The only remaining source of background is the production of the 
five-pion final state
($e^+e^-\to\omega\pi^+\pi^-$  and $e^+e^-\to\eta\pi^+\pi^-$ with
$\omega$ and $\eta$ decays to $\pi^+\pi^-\pi^0$), which results in 
a lower total energy because of a missing $\pi^0$. The contributions of
these channels are shown  in Fig.~\ref{distr2}(a)  by the 
hatched and cross-hatched histograms, respectively, and the 
histograms were obtained from simulation and the values of the 
corresponding cross sections measured at CMD-2~\cite{CMD2_5pi}. The
applied cut E$_{\rm tot}>$~2E$_{\rm beam}-130$~MeV shown by the 
vertical line  almost completely  removes these
events. The missing mass squared distribution after this cut is shown in
Fig.~\ref{distr2}(b) in comparison with simulation.
For four-track events the remaining background is estimated to be
less than 1\%.

\hspace*{\parindent}
To increase the data sample and improve a systematic uncertainty
related to the efficiency determination, a sample of events with three 
selected tracks was additionally used to determine the number of 
four-pion events with one missing track.
A track can be lost for one of the following reasons: it 
flies at  small polar angles outside the 
efficient DC region, decays in flight, because of incorrect
reconstruction,  due to nuclear interactions, by overlapping with 
another track. From energy-momentum conservation the direction and 
momentum of a missing track can be calculated assuming a four-pion 
final state. The reconstructed momentum vectors were used to apply 
the additional requirements on the angles between two tracks in 
the R-$\phi$ plane described above. 

\begin{center}
\begin{figure}[tbh]
\vspace{-0.2cm}
\psfig{file=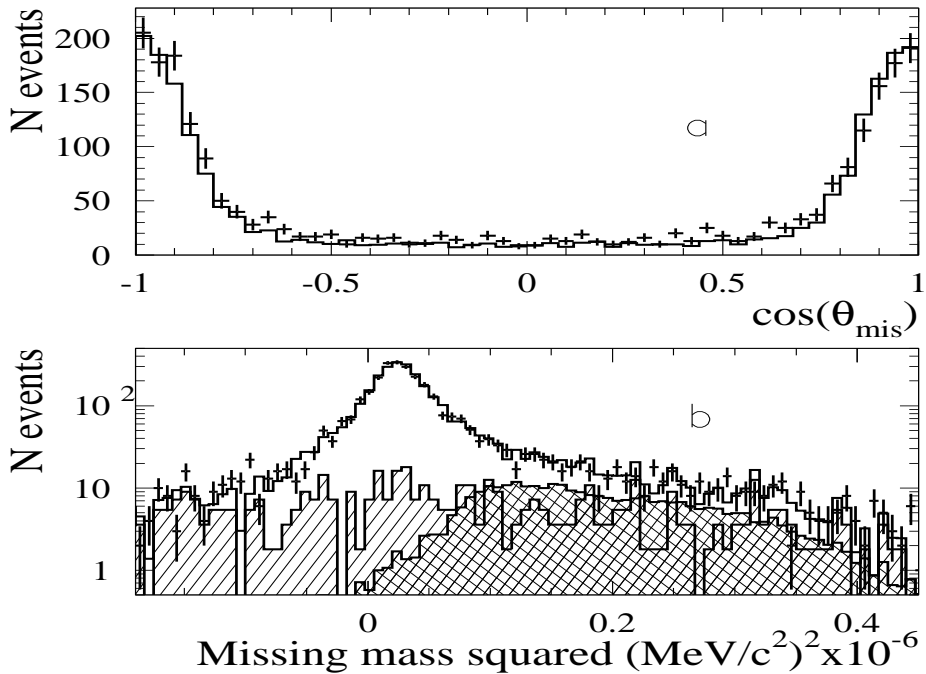, 
width=0.8\textwidth, height=0.6\textwidth}
\caption
{
Distributions for three-track experimental events (points with errors):
(a) Cosine of a polar angle for a missing pion, the open histogram shows 
results of the simulation;
(b) Missing mass squared distribution for the three-track sample.
The open histogram shows results of the simulation.
The hatched histogram shows a contribution from beam-gas
background. The cross-hatched histogram shows a contribution from five-pion
events.
}
\label{distr3}
\end{figure}
\end{center}

Figure~\ref{distr3}(a) shows the $cos(\theta)$
distribution for a missing pion after the cuts
on the angles between two tracks. It can be
seen  that most of the three-track events have a fourth track 
flying at small angles with respect to the beam axis and not detected by 
the DC. In some cases the  missing track is inside the DC acceptance 
but does not meet the selection criteria. 

The three-track event sample has higher background than the four-track
one, but events corresponding to the four-pion final state could be 
separated by requiring that a missing particle have the charged pion
mass. The distribution of  missing mass  squared for three-track events
is shown in Fig.~\ref{distr3}(b) and
exhibits a clear signal at the pion mass that can be
attributed to four-pion events. 
The background in this sample comes mostly from five-pion events 
and beam-gas interactions. 
The latter source results in a relatively flat distribution over missing
mass and can be estimated from the events in the 7.0 $<|Z_{}|<$ 10.0
cm region (the longitudinal size of the collision region has r.m.s. about 
2 cm). This contribution is shown by the hatched histogram in 
Fig.~\ref{distr3}(b).
The contribution from five-pion events estimated from the MC simulation
is shown by the cross-hatched histogram in Fig.~\ref{distr3}(b).

To obtain the number of four-pion events from a three-track sample,
the distribution shown in Fig.~\ref{distr3}(b) was fit with a sum of
functions describing a pion peak and background.
The pion peak line shape was taken from simulation 
of the four-pion process and had a Gaussian shape with a small 
radiative tail. All parameters of this function were fixed except 
for the number of events. A second order polynomial with all free 
parameters was used for background.
As a result of the fit, the number of four-pion events in the 
three-track sample was determined.



To check stability of the background subtraction procedure, 
the number of four-pion events was independently estimated by 
statistical subtraction of background shown in Fig.~\ref{distr3}(b). 
This procedure gives results consistent with those from the fit,
but has slightly higher errors in the number of four-pion events.
The 2\% variation in the number of events between the two subtraction
procedures was taken as an estimate of a systematic error.

About 20550 four-track and 17180 three-track events survive at this
stage of selection. The number of four- and three-track events determined 
at each energy is listed in Table~\ref{table}.

\begin{table}[tb]

\caption{Luminosity, number of events, detection efficiency, 
rad. correction, cross section and vacuum polarization correction}
\label{table}
\smallskip
\begin{center}
\renewcommand{\arraystretch}{0.95}
\begin{tabular}{cccccccc}
\hline
\hline
{2E$_{\rm beam}$, MeV}&{L,nb$^{-1}$}&{\bf $N_{4tr}$}&{\bf $N_{3tr}$}
&{\bf $\epsilon_{MC}$ } &   {\bf $1+\delta$ }&{$\sigma$, nb}& 
$|1-\Pi({\rm 2E}_{\rm beam})|^2$ \\
\hline
 980.0&     59.00&      7&      7.32$\pm$ 4.08&     0.426&     0.880&      0.65$\pm$0.22& 0.9751\\
1040.0&     72.76&     25&     32.00$\pm$ 8.05&     0.442&     0.881&      2.01$\pm$0.33& 0.9583\\
1050.0&    116.03&     44&    47.65$\pm$  7.99&     0.444&     0.882&      2.02$\pm$0.23& 0.9622\\
1060.0&     75.71&     40&    35.15$\pm$  7.03&     0.447&     0.882&      2.52$\pm$0.32& 0.9643\\
1070.0&     80.63&     43&    36.16$\pm$  7.04&     0.449&     0.882&      2.48$\pm$0.30& 0.9657\\
1080.0&     59.00&    37&     32.77$\pm$  7.03&     0.452&     0.883&      2.96$\pm$0.40& 0.9666\\
1090.0&     83.58&     58&    60.81$\pm$  8.80&     0.454&     0.883&      3.55$\pm$0.35& 0.9674\\
1100.0&     57.03&     50&    43.56$\pm$  7.51&     0.456&     0.883&      4.07$\pm$0.45& 0.9679\\
1110.0&     82.60&     62&    69.30$\pm$  9.44&     0.458&     0.885&      3.92$\pm$0.37& 0.9684\\
1120.0&     58.01&     37&    54.00$\pm$  9.70&     0.460&     0.885&      3.85$\pm$0.48& 0.9688\\
1130.0&     96.36&    101&    90.82$\pm$  10.70&     0.462&     0.886&     4.86$\pm$0.37& 0.9692\\
1140.0&     96.36&    129&   110.85$\pm$  11.57&     0.464&     0.885&     6.05$\pm$0.41& 0.9695\\
1150.0&     52.11&     66&    52.73$\pm$   8.29&     0.466&     0.887&     5.51$\pm$0.54& 0.9697\\
1160.0&    111.11&    155&   135.56$\pm$  12.64&     0.468&     0.886&     6.30$\pm$0.38& 0.9700\\
1170.0&     90.46&    136&   130.49$\pm$  17.91&     0.470&     0.888&     7.05$\pm$0.57& 0.9702\\
1180.0&    113.08&    204&   177.41$\pm$  14.87&     0.472&     0.889&     8.04$\pm$0.43& 0.9704\\
1190.0&    128.81&    256&   246.67$\pm$  18.12&     0.474&     0.889&     9.27$\pm$0.45& 0.9706\\
1200.0&    183.87&    417&   339.41$\pm$  19.96&     0.475&     0.890&     9.73$\pm$0.37& 0.9708\\
1210.0&    120.94&    295&   215.23$\pm$  16.56&     0.477&     0.892&     9.92$\pm$0.46& 0.9711\\
1220.0&    111.11&    323&   229.16$\pm$  16.78&     0.479&     0.892&     11.65$\pm$0.52& 0.9712\\
1230.0&    140.61&    422&   332.33$\pm$  19.81&     0.480&     0.891&     12.54$\pm$0.47& 0.9714\\
1240.0&    141.59&    460&   314.80$\pm$  19.51&     0.481&     0.893&     12.73$\pm$0.48& 0.9716\\
1250.0&    208.46&    627&   579.42$\pm$  31.68&     0.483&     0.894&     13.41$\pm$0.45& 0.9717\\
1260.0&    176.99&    677&   492.77$\pm$  23.78&     0.484&     0.894&     15.26$\pm$0.46& 0.9719\\
1270.0&    242.87&    912&   751.12$\pm$  29.63&     0.485&     0.896&     15.75$\pm$0.40& 0.9721\\
1280.0&    219.27&    859&   684.76$\pm$  29.13&     0.487&     0.896&     16.14$\pm$0.43& 0.9723\\
1290.0&    285.15&   1067&  1011.61$\pm$  41.13&     0.488&     0.898&     16.64$\pm$0.42& 0.9724\\
1300.0&    279.09&   1256&   949.54$\pm$  33.03&     0.489&     0.898&     17.99$\pm$0.40& 0.9726\\
1310.0&    231.07&   1055&   829.30$\pm$  31.40&     0.490&     0.899&     18.50$\pm$0.44& 0.9728\\
1320.0&    191.74&    893&   779.53$\pm$  30.26&     0.491&     0.901&     19.72$\pm$0.50& 0.9730\\
1330.0&    320.66&   1377&  1337.15$\pm$  43.74&     0.492&     0.902&     19.07$\pm$0.40& 0.9732\\
1340.0&    204.52&   1170&   754.22$\pm$  30.03&     0.493&     0.904&     21.12$\pm$0.50& 0.9734\\
1350.0&    229.11&   1085&  1101.86$\pm$  41.34&     0.494&     0.904&     21.37$\pm$0.52& 0.9737\\
1360.0&    313.67&   1615&  1499.79$\pm$  45.99&     0.495&     0.906&     22.16$\pm$0.43& 0.9739\\
1370.0&    186.82&   1023&   935.66$\pm$  36.41&     0.495&     0.907&     23.35$\pm$0.58& 0.9742\\
1380.0&    575.22&   3568&  2617.01$\pm$  68.22&     0.496&     0.909&     23.86$\pm$0.35& 0.9746\\
\hline
\end{tabular}
\end{center}
\end{table}

\section*{Detection efficiency from simulation}
\begin{center}
\begin{figure}[tbh]
\vspace{-0.2cm}
\psfig{file=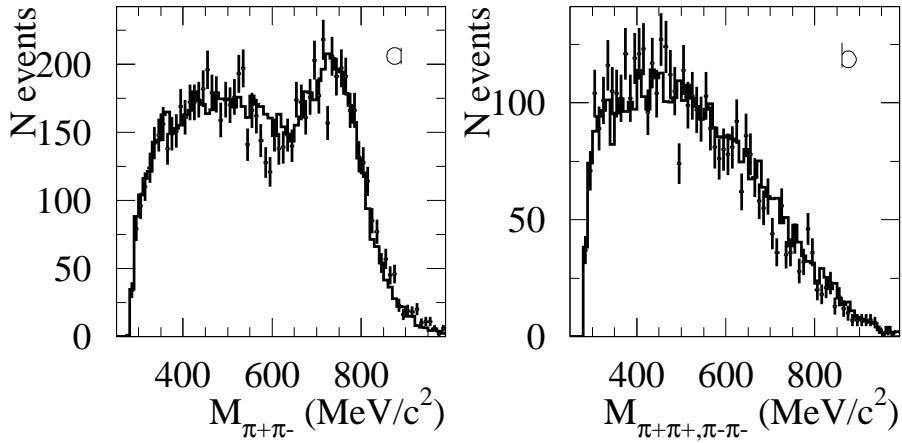, 
width=0.8\textwidth, height=0.4\textwidth}
\caption
{
Di-pion mass distributions for experimental events  (points with errors)
and simulation (histograms):
(a) $\pi^+\pi^-$ mass spectrum;
(b) $\pi^{\pm}\pi^{\pm}$ mass spectrum
}
\label{distr4}
\end{figure}
\end{center}

\hspace*{\parindent}
The detailed study of the process dynamics performed 
in~\cite{CMD2_4pi} showed that the $a_{1}(1260)\pi$ intermediate
mechanism dominates the final state with four charged pions.
Various observed distributions were investigated in that analysis to 
search for a possible admixture of some other mechanisms 
like $\rho f_0(600), a_2(1320)\pi, \pi(1300)\pi$ etc. It was shown that
the $a_{1}(1260)\pi$ by itself can account for
the observed spectra although a small admixture of other mechanisms
can not be excluded. The highest upper limit for a possible
admixture, equal to 15\%,  was obtained for the 
$\pi(1300)\pi$ model. Therefore,  for Monte Carlo simulation and 
studies of the detection efficiency (acceptance) we used the
$a_{1}(1260)\pi$ and $\pi(1300)\pi$ models described in~\cite{CMD2_4pi}. 
The two-pion invariant mass experimental spectra shown
in Fig.~\ref{distr4} together with
those from the simulation within the  $a_{1}(1260)\pi$ model 
at 2E$_{\rm beam}=$~1380~MeV qualitatively agree with each other.
The model with a 15\% admixture of the  $\pi(1300)\pi$ mechanism
gives mass distributions indistinguishable from those for the
$a_{1}(1260)\pi$ model. 
%
%


\label{ratio}

The detection efficiency was determined from MC simulation using
both four- and three-track events. 
It should be pointed out that in this 
case possible data-MC inconsistencies in the description of the DC 
inefficiency and (partly) in the model-dependent angular distributions 
are compensated, because in case of an undetected track 
an event migrates from the four- to the three-track sample. 
MC simulation was performed at nine energy points from 
980 to 1380~MeV and 30000 events were generated at each of these 
points. The detection efficiency thus obtained  monotonously grows 
from 42.6\% to 49.6\%  in the energy range studied. Its
values at each energy point shown in 
Table~\ref{table} were calculated using the polynomial approximation 
of the detection efficiency determined at the nine points 
mentioned above. Its statistical error is less than 1\%.


\section*{Cross Section Calculation}    
\hspace*{\parindent}
At each energy the cross section was calculated as 
$$
\sigma = \frac{N_{4 \rm tr}+N_{3 \rm tr}}{L\cdot\epsilon\cdot(1+\delta)},
$$ 
where 
$L$ is the integrated luminosity for this energy point, $\epsilon$ is 
the detection efficiency obtained from the MC simulation 
and $(1+\delta)$ is the radiative correction calculated 
according to~\cite{kur_fad}. The charged trigger efficiency
was studied in Ref.~\cite{kskl} where it was shown that for two tracks
the trigger efficiency was $(98.3 \pm 0.9 \pm 0.5)\%$.  
Since only one charged track is 
sufficient for a trigger, we assume 
that for multitrack events considered in this analysis the 
trigger efficiency is close to 100\%. 

The integrated luminosity, the number of four and three-track events,
detection efficiency, radiative correction  and obtained cross section
for each energy point are listed in Table~\ref{table}. 
Table~\ref{table} also contains a so called vacuum polarization 
correction factor $|1-\Pi({\rm 2E}_{\rm beam})|^2$, where 
$\Pi({\rm 2E}_{\rm beam})$ is the polarization operator. Multiplying 
it by the ``dressed'' cross section presented in Table~\ref{table}, 
one obtains the ``bare'' cross section to be used in dispersion 
integral calculations (see, e.g., the discussion in Ref.~\cite{update}). 
\section*{Systematic errors}
\hspace*{\parindent}
The following sources of systematic uncertainties were considered.

\begin{itemize}
\item {
The model dependence of the acceptance is determined by the angular
distribution, which is specific for each particular model. Therefore,
we compared results of the cross section calculation for different 
$cos(\theta)$ cuts in two models of the final state production: 
the dominant $a_1(1260)\pi$ mechanism and $\pi(1300)\pi$, which 
admixture at the 15\% level, as discussed above, was not excluded 
by the analysis in~\cite{CMD2_4pi}. The resulting systematic 
uncertainty caused by model and angular cut dependence is 
estimated as 3\%.
}  
\item{
A systematic error because of the selection criteria other than 
the angular cuts was studied by 
varying the cuts described previously and doesn't exceed 2\%. 
}
\item{
The uncertainty in the determination of the integrated luminosity 
comes from the selection criteria of Bhabha events, radiative
corrections and calibrations of DC and CsI and does not exceed 
2\%~\cite{prep}.
}
\item
The contribution of the uncertainty of the charged trigger inefficiency 
studied with $\phi \to K^0_{\rm S}K^0_{\rm L}$ events~\cite{kskl}
appears to be much less than 1\% and can be neglected.
\item{
A possible uncertainty in the beam energy was studied using the momentum
distribution of Bhabha events and total energy of four-pion
events. The uncertainty at the level of 10$^{-3}$ was not excluded and
because of the relatively fast cross section variation it can
result in a 1\% change of the cross section.}
\item{
A radiative correction uncertainty was estimated as about 
1\% mainly due to the uncertainty in the maximum allowed energy of the 
emitted photon at the integration of the formulae 
from~\cite{kur_fad} as well as the accuracy of these formulae.
}
\item{
The uncertainty because of background subtraction for four-track 
(three-track) events is estimated as 1\% (2\%) above 1100~MeV 
growing to 5\% below 1100~MeV for both types of events. 
At low energy the cross sections of the processes 
$e^+e^- \to \pi^+\pi^-\pi^0$ and  $e^+e^- \to \pi^+\pi^-2\pi^0$
dominating the background are considerably higher than that of the 
process under study.
}
\end{itemize}


The above systematic uncertainties summed in quadrature give an overall
systematic error of about 5\% above 1100~MeV and about 7\% below 
1100~MeV. This uncertainty is common (energy-independent) for most
of the energy range studied. Some energy-dependent contribution to
the total experimental uncertainty is possible below 1100~MeV, but 
there the systematic error is much smaller than a statistical one.

\section*{Discussion}
\hspace*{\parindent}

\begin{center}
\begin{figure}[tbh]
\vspace{-0.2cm}
\psfig{file=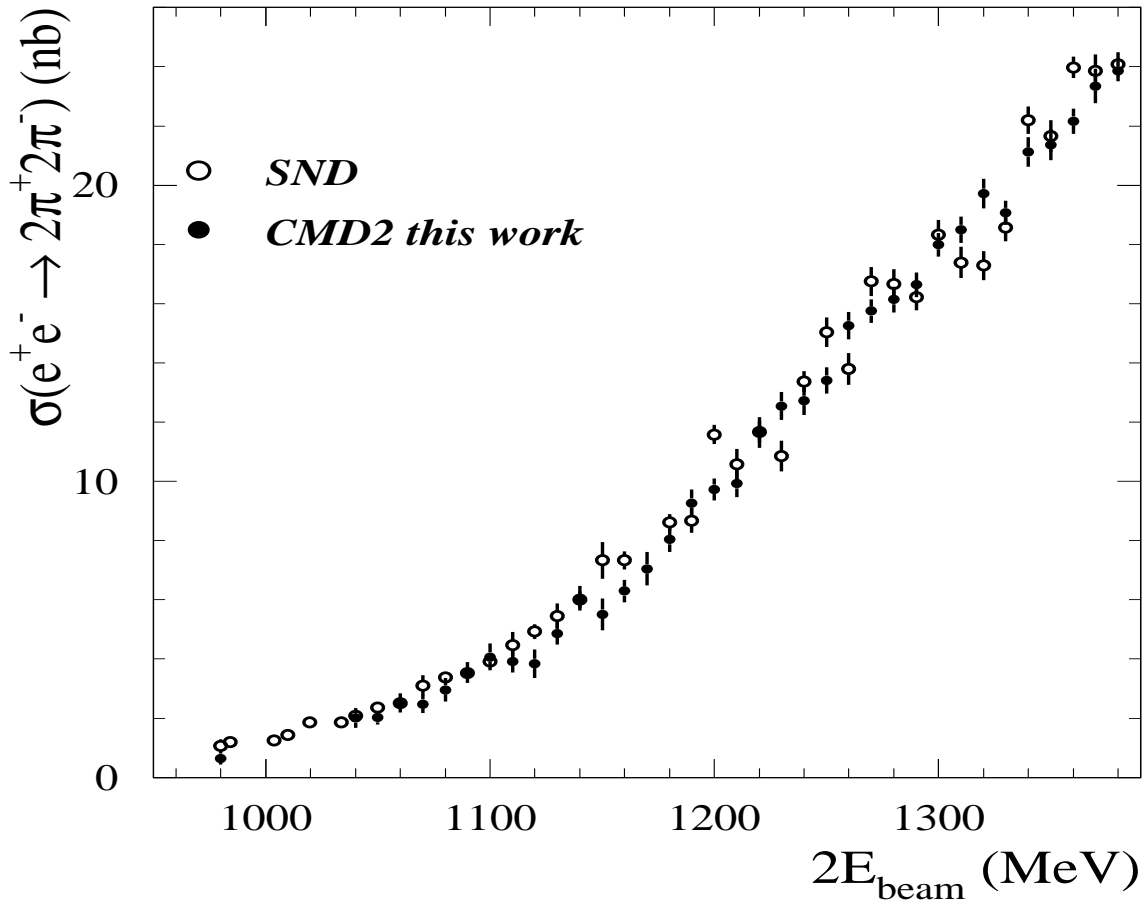, 
width=1.0\textwidth, height=0.8\textwidth}
\vspace{-0.8cm}
\caption
{
Cross section of the process $e^+e^-\to\pi^+\pi^-\pi^+\pi^-$ obtained in
two most precise experiments in the 900-1400 MeV energy range. Only 
statistical errors are shown.
}
\label{4pi_all}
\end{figure}
\end{center}

From Fig.~\ref{4pi_all} it is clear that the obtained values of the
cross section are consistent with  the results of the precise
measurement performed by the SND group~\cite{SND_4pi2}.
They are also in good
agreement with the other previous experiments in the energy 
range studied~\cite{olya,cmd,NDR}.
The  three-track events used in this analysis allowed a significant
increase of the data sample and improvement of the systematic
uncertainties. 
 

The rapid growth of the cross section with energy is apparently due to 
the $\rho(1450)$ and $\rho(1700)$. The maximum energy of our experiment
is insufficient for a quantitative study of these resonances which 
parameters are currently known with rather bad precision~\cite{pdg}.
We hope that future experiments at the VEPP-2000 collider currently
under construction in Novosibirsk~\cite{v2000} will allow a detailed 
investigation of both $4\pi$ final states from threshold to 2000~MeV. 
  
Let us estimate the implication of our results for the corresponding
contribution to $a_{\mu}^{\rm had,LO}$, the leading order hadronic term in
the muon anomalous magnetic moment. 
To this end we calculate its value in the c.m.energy range studied in
this work (from 1040 to 1380~MeV) using recent precise data from
SND and CMD-2 and compare it to that  based on the 
previous $e^+e^-$ measurements~\cite{olya,cmd,NDR} in Table~\ref{g-2}. 

The first line of the Table (Old data) gives the result based 
on the data of OLYA, CMD and ND while the second one (New data) 
is obtained from the recent data of SND and CMD-2, which are in 
good agreement with each other:
$(4.36 \pm 0.31) \cdot 10^{-10}$~(SND) vs.
$(4.24 \pm 0.20) \cdot 10^{-10}$~(CMD-2).
The third line (Old + New) presents the weighted average of   
these two estimates. 
For convenience, we list separately statistical and systematic uncertainties
in the second column while the third one gives the total error 
obtained by adding them in quadrature. One can see that the 
estimate based on the new data is in good agreement with that 
coming from the old data. Because of the large number of energy points
in all the measurements at VEPP-2M, an overall statistical error is 
much smaller than a corresponding systematic uncertainty for both 
old and new data. 
The statistical precision of the new measurements with SND and CMD-2 is 
three times higher than before and the total error is almost a factor
of two smaller than earlier.
The combined value based on both old and new data is 
dominated by the new measurements and provides a significant improvement 
of the accuracy in the $2\pi^+2\pi^-$ contribution to 
$a_{\mu}^{\rm had.LO}$. 
Although the relative contribution to $a^{\rm had,LO}_{\mu}$ of the 
considered channel and the energy range from 980 to 1380~MeV is 
small, in combination with the 
$\pi^+\pi^-2\pi^0$ final state the 4$\pi$ production is responsible for 
about 57\% of the contribution to  $a^{\rm had,LO}_{\mu}$ from
the hadronic continuum below 2000~MeV, i.e. production of hadrons 
not from the $\rho$, $\omega$ and $\phi$. Therefore, significant 
improvement of the precision of its cross section is of
extreme importance for the interpretation of the current and future
measurements of the muon anomalous magnetic moment~\cite{bnl}.


 
\begin{table}
\caption{\label{g-2} Contributions of the $2\pi^+2\pi^-$ channel 
to $(g_{\mu}-2)/2$}
\begin{tabular}{lcc}
\hline
Data  & a$_{\mu}^{\rm had,LO}$, 10$^{-10}$ & Total error, 10$^{-10}$ \\
\hline
Old   & 4.40 $\pm$ 0.06 $\pm$ 0.31 & 0.31 \\
\hline
New & 4.28 $\pm$ 0.02  $\pm$ 0.17 & 0.17 \\
\hline
Old + New & 4.31 $\pm$ 0.02 $\pm$ 0.15 & 0.15 \\ 
\hline
\end{tabular}
\end{table}    


\section*{ \boldmath Conclusion}
\hspace*{\parindent}
The total cross section of the process $e^+e^-\to\pi^+\pi^-\pi^+\pi^-$ 
has been measured using 5.8 pb$^{-1}$ of integrated luminosity 
collected with the CMD-2 detector at the VEPP-2M $e^+e^-$ collider. 
The new refined analysis based on the extraction of the detector efficiency
from three- and four-track event samples results in a factor of two
larger data sample and allows reduction of systematic
errors. 
The observed production mechanism is consistent 
with the $a_1(1260)\pi$ intermediate state. The values of the 
obtained cross section are in good agreement with all
other experiments in the energy range studied and supersede
our previous results based on the same data sample~\cite{CMD2_4pi}.

%
\subsection*{Acknowledgments}
\hspace*{\parindent}
The authors are grateful to V.P.~Druzhinin and Z.K.~Silagadze
for useful discussions.
This work is supported in part by grants
DOE DEFG0291ER40646, NSF PHY-9722600, NSF PHY-0100468,
PST.CLG.980342, RFBR-02-02-16126-a, RFBR-03-02-10843 and 
RFBR-04-02-16223-a.

\end{document}